\documentclass{ws-procs10x7}

\begin{document}

\title{Isospin Amplitude Analysis of $B \to \pi\pi$ 
       Decays \\ within the Standard Model} 

\author{A.Ya. Parkhomenko}

\address{
         University of Bern, Sidlerstrasse 5, 
         CH-3012 Bern, Switzerland \\
%
          Yaroslavl State University, Sovietskaya 14,
          150000 Yaroslavl, Russia \\  
          E-mail: parkh@uniyar.ac.ru}  

\twocolumn[\maketitle\abstract{ 
Taking into account the recent measurements of the branching ratios and 
CP asymmetries in the $B \to \pi\pi$ decays by the BABAR and BELLE 
collaborations, an amplitude analysis based on the isospin symmetry 
is reported. These data allow to get an independent determination 
of the CKM unitarity-triangle angle~$\gamma$. One of the best-fit 
solutions gives the range $\gamma = (65 \pm 5)^\circ$, in 
excellent agreement with the global CKM fit within the Standard Model. 
The other parameters from this fit confirm the earlier analyses that  
the data on $B \to \pi\pi$ decays  require various topological 
amplitudes (neglecting the electroweak ones) to be of comparable 
magnitude with large strong-phase differences. 
}]

\section{Introduction}
\label{sec:introduction}

Recent experimental investigations of  exclusive $B$-meson decays  
provide a quantitative test of theoretical frameworks such 
as the QCD-Factorization\cite{Beneke:1999br,Beneke:2003zv}, 
perturbative-QCD (pQCD)\cite{Keum:2000ph} and Soft-Collinear 
Effective Theory (SCET)\cite{Bauer:2004tj,Bauer:2004ck}. 
Data on the $B \to \pi\pi$ decays obtained by the CLEO, 
BABAR and BELLE collaborations have been averaged by 
HFAG\cite{HFAG} and are displayed in Table~\ref{tab:exp-data}. 
We perform a phenomenological analysis of this data with the 
assumption of the isospin symmetry among the decay amplitudes.
There are several ways to go about it.  
A possibility is to assume that the CKM unitarity-triangle 
(UT) angles~$\beta$ and~$\gamma$ are known from the global CKM 
fit of the Standard Model (SM) or from direct measurements
elsewhere (for example, $\sin(2\beta)$ from $B \to J/\psi K^{(*)}$ 
and~$\gamma$ from the $B \to \pi K$ decays). In this case, the 
isospin amplitude analysis allows to fix different topological 
contributions (color-allowed tree~$T$, color-suppressed tree~$C$, 
penguin~$P$ and electroweak-penguin~$P_{\rm EW}$) originated 
by strong interactions\cite{Charng:2004ed}. Alternatively, 
electroweak penguins in $B \to \pi\pi$ decays are assumed 
to be small and neglected, and the amplitudes are parameterized 
in terms of tree and penguin contributions only, enabling 
an independent complete amplitude analysis of the data 
(including the determination of~$\gamma$). This is the 
generalization of our previous analysis\cite{Ali:2004hb} 
based on the $B^0 \to \pi^+\pi^-$ data alone and is the 
focus of this report. 

\begin{table}[tb]
\caption{Current experimental data on the $B \to \pi\pi$ 
         decays averaged by the HFAG$^6$. Branching
         ratios are in units of $10^{-6}$.} 
\label{tab:exp-data} 
\begin{center} 
\begin{tabular}{|l|c|}
\hline 
Quantity & Average 
\\ \hline  
$\bar {\cal B} (B^+ \to \pi^+ \pi^0)$ & 
$5.5 \pm 0.6$ \\
$\bar {\cal B} (B^0 \to \pi^+ \pi^-)$ & 
$4.6 \pm 0.4$ \\
$\bar {\cal B} (B^0 \to \pi^0 \pi^0)$ & 
$1.51 \pm 0.28$ \\ \hline 
$A_{\rm CP} (\pi^+ \pi^0)$ & 
$-0.02 \pm 0.07$ \\ 
$A_{\rm CP} (\pi^+ \pi^-) = - C_{\pi\pi}^{+-}$ & 
$0.37 \pm 0.11$ \\ 
$S_{\pi\pi}^{+-}$ & 
$-0.61 \pm 0.14$ \\ 
$A_{\rm CP} (\pi^0 \pi^0)= - C_{\pi\pi}^{00}$ & 
$0.28 \pm 0.39$ 
\\ \hline  
\end{tabular} 
\end{center}
\end{table}
\begin{table*}[htb]
\caption{The best-fit solutions of the $\chi^2$-function 
         for the parameters entering the $B \to \pi\pi$ 
         decay amplitudes. The topological contributions 
         $T_c$, $P_c$ and $C_c$ are expressed in units  
         of $10^{-3} N_\pi$, where $N_\pi = (G_F/\sqrt 2) \, 
         m_B^2 \, f_\pi \simeq 3.0 \times 10^{-5} \, {\rm GeV}$.}  
\label{tab:fit-results}
\begin{center} 
\begin{tabular}{|c|cc|cc|} 
\hline  
& I & II & III & IV  
\\ \hline 
$\gamma$ & 
$(65.3^{+4.7}_{-5.2})^\circ$ & $( 27.1^{+2.9}_{-3.1})^\circ$ & 
$(16.3^{+3.9}_{-4.0})^\circ$ & $(158.2^{+5.0}_{-4.7})^\circ$  
\\[1mm] 
$r_c$ & 
$0.51^{+0.10}_{-0.09}$ & $0.51^{+0.03}_{-0.03}$ & 
$1.89^{+0.08}_{-0.07}$ & $1.89^{+0.10}_{-0.09}$  
\\[1mm] 
$\delta_c$ & 
$(- 39.4^{+10.3}_{-9.8})^\circ$ & $(-161.4^{+6.5}_{-5.7})^\circ$ & 
$(-133.1^{+ 5.9}_{-4.8})^\circ$ & $(- 75.1^{+6.3}_{-8.0})^\circ$  
\\[1mm] 
$x_c$ & 
$1.11^{+0.09}_{-0.10}$ & $0.16^{+0.13}_{-0.14}$ & 
$2.82^{+0.10}_{-0.11}$ & $3.10^{+0.13}_{-0.14}$  
\\[1mm]    
$\Delta_c$ & 
$(- 55.7^{+13.5}_{-12.5})^\circ$ & $(116.9^{+17.5}_{-15.9})^\circ$ & 
$(-140.4^{+ 6.8}_{- 6.2})^\circ$ & $(113.7^{+ 7.5}_{- 8.1})^\circ$  
\\[1mm] \hline
$|T_c|$ & 
$0.60^{+0.03}_{-0.03}$ & $1.20^{+0.03}_{-0.03}$ & 
$0.52^{+0.11}_{-0.09}$ & $0.40^{+0.07}_{-0.08}$  
\\[1mm]
$|P_c|$ & 
$0.31^{+0.04}_{-0.04}$ & $0.61^{+0.03}_{-0.03}$ & 
$1.00^{+0.03}_{-0.03}$ & $0.75^{+0.03}_{-0.03}$  
\\[1mm] 
$|C_c|$ & 
$0.67^{+0.05}_{-0.06}$ & $0.19^{+0.16}_{-0.16}$ & 
$1.48^{+0.05}_{-0.06}$ & $1.23^{+0.05}_{-0.06}$  
\\[1mm] \hline   
$S_{\pi\pi}^{00}$ & 
$ 0.73^{+0.16}_{-0.22}$ & $-0.72^{+0.07}_{-0.06}$ & 
$-0.72^{+0.20}_{-0.22}$ & $ 0.73^{+0.32}_{-0.18}$  
\\ \hline 
\end{tabular}
\end{center}
\end{table*}

\section{Theoretical Input}
\label{sec:th-input}

The parameterization of the $\bar B \to \pi \pi$ decay 
amplitudes in the $c$-convention is as follows: 
{\small 
\begin{eqnarray} 
&& \hspace*{2mm}
\bar {\cal A}^{+-} \equiv  
- |T_c| \, {\rm e}^{i \delta_c^T} \, {\rm e}^{- i \gamma} 
\left [ 1 + r_c \, {\rm e}^{i \delta_c}  
{\rm e}^{i \gamma} \right ] , 
\nonumber \\ 
&& \hspace*{-2mm}
\sqrt 2 \, \bar {\cal A}^{-0} \equiv  
- |T_c| \, {\rm e}^{i \delta_c^T} \, {\rm e}^{- i \gamma} 
\left [ 1 + x_c \, {\rm e}^{i \Delta_c} \right ] , 
\label{eq:GR-parameterization} \\ 
&& \hspace*{-2mm}
\sqrt 2 \, \bar {\cal A}^{00} \equiv  
- |T_c| \, {\rm e}^{i \delta_c^T} \, {\rm e}^{- i \gamma} 
\left [ x_c \, {\rm e}^{i \Delta_c} - 
r_c \, {\rm e}^{i \delta_c} {\rm e}^{i \gamma} \right ] , 
\nonumber 
\end{eqnarray}
}%
where $r_c = |P_c/T_c|$ and $\delta_c$ are the magnitude 
and phase of the penguin-to-tree ratio, $x_c = |C_c/T_c|$ 
and~$\Delta_c$ are the magnitude and phase of the 
color-suppressed to color-allowed amplitude ratio, respectively.  
These amplitudes satisfy the isospin relation: 
\begin{equation} 
\bar {\cal A}^{+-} + \sqrt 2 \, \bar {\cal A}^{00} = 
\sqrt 2 \, \bar {\cal A}^{-0}.  
\label{eq:isospin-rel}
\end{equation} 
The same equation can be written for the charged-conjugate 
amplitudes ${\cal A}^{+-}$, ${\cal A}^{00}$ and~${\cal A}^{+0}$.  

The charged-conjugate averaged branching ratios are 
related with the decay amplitudes as follows: 
{\small 
\begin{equation} 
\hspace*{-2mm}
\bar{\cal B} (B \to \pi^i \pi^j) = 
\frac{\tau_B}{32 \pi m_B} 
\left [ |{\cal A}^{ij}|^2 + |\bar {\cal A}^{ij}|^2 \right ] , 
\label{eq:branch-def}
\end{equation}
}%
where the amplitudes have the dimension of a mass.
%
%
Explicit expressions for the direct~$C_{\pi\pi}^{+-}$ 
and mixing-induced~$S_{\pi\pi}^{+-}$ CP asymmetries in the 
$B^0 \to \pi^+ \pi^-$ decays can be found in\cite{Ali:2004hb} and 
the same definitions for~$C_{\pi\pi}^{00}$ and~$S_{\pi\pi}^{00}$ 
are adopted for the $B^0 \to \pi^0 \pi^0$ decay. In the limit of 
neglecting penguin contribution ($P_c = 0$) in both decay modes,  
the CP asymmetries simplify greatly, yielding 
$C_{\pi\pi}^{+-} = C_{\pi\pi}^{00} = 0$ and  
$S_{\pi\pi}^{+-} = S_{\pi\pi}^{00} = \sin (2\alpha)$, 
where the CKM unitarity-triangle relation 
$\alpha + \beta + \gamma = \pi$ is used.

\section{Results of the Fit}
\label{sec:fit} 

The results of the fits within the physical ranges 
of the parameters: $r_c > 0$, $|\delta_c| < 180^\circ$, 
$x_c > 0$, $|\Delta_c| < 180^\circ$, and $0 < \gamma < 180^\circ$ 
at the fixed value $\beta = 23.3^\circ$ 
are presented in Table~\ref{tab:fit-results}, yielding four 
minimal points shown as fits~I to~IV.
The last quantity presented in Table~\ref{tab:fit-results} 
is the mixing-induced CP asymmetry~$S_{\pi\pi}^{00}$ in the 
$B^0 \to \pi^0 \pi^0$ decay. Note that $|S_{\pi\pi}^{00}|$ 
is numerically very close to the experimental world average 
$\sin(2\beta) = 0.726 \pm 0.037$ resulting from the 
analysis\cite{HFAG} of the time-dependent CP asymmetry in the 
$B \to J/\psi K^{(*)}$ decays. The $1\sigma$ intervals of a specific 
fitted parameter are obtained as the solutions of the equation 
$\chi^2 = \chi^2_{\rm min} + 1$ under the condition that all 
the fitted parameters are kept at their central values except 
the considered one.   

\begin{figure}[bt] 
\includegraphics[width=0.45\textwidth]{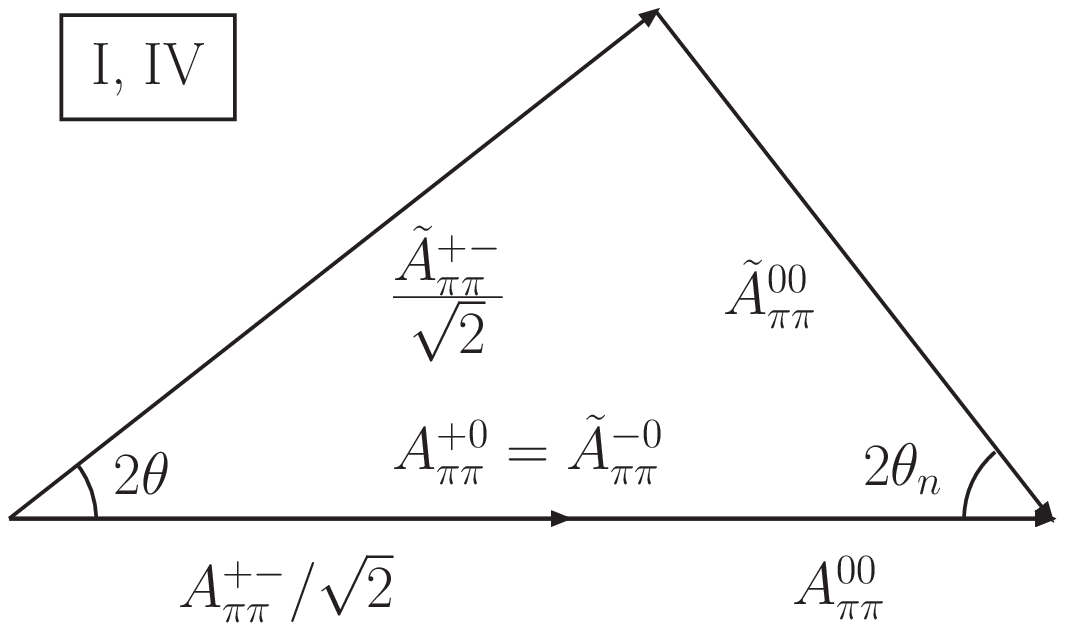} 
\newline 
\vspace*{5mm} 
\includegraphics[width=0.45\textwidth]{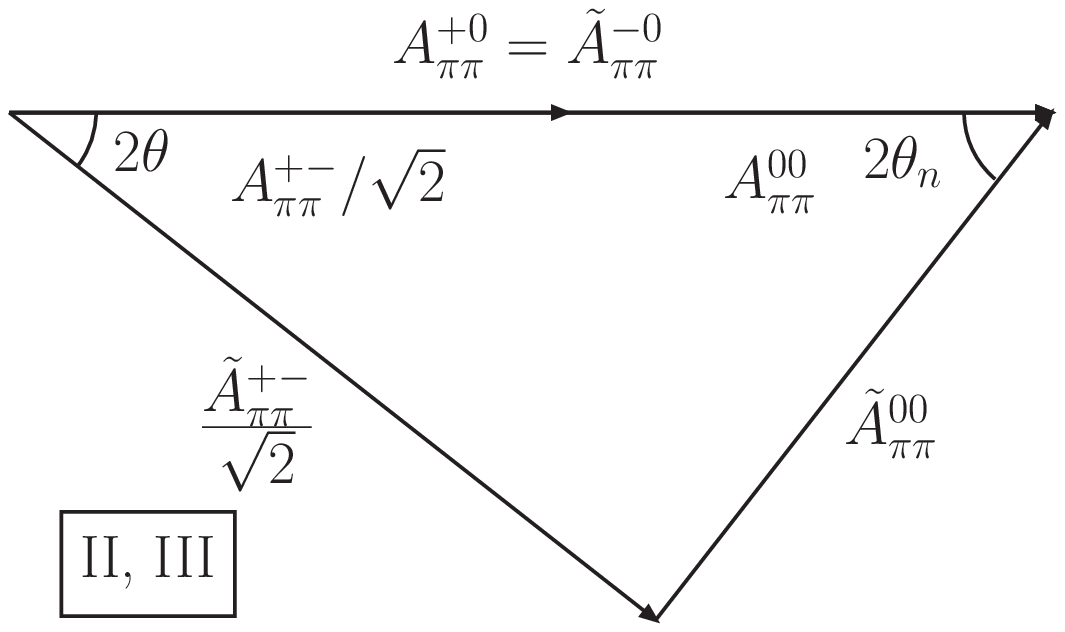}  
%
\caption{The graphical representation of the best-fit solutions
         in the complex space of the $B \to \pi\pi$ decay amplitudes. 
         The strong phases of the $B^+ \to \pi^+\pi^0$ 
         amplitude~${\cal A}^{+0}$ and the phase-shifted  
         charged-conjugate one $\tilde{\cal A}^{-0} = 
         {\rm e}^{+2i\gamma} \, \bar{\cal A}^{-0}$ are chosen in 
         such a way that both amplitudes are real and positive.}
\label{fig:isospin}
\end{figure}
%

Solution~IV contains the extremely large value for 
the angle~$\gamma$ which is in obvious contradiction 
with the global CKM fit value $\gamma = (65 \pm 7)^\circ$ 
obtained within the SM\cite{Ali:2004hb}. 
Small values for this angle are found in the best-fit 
solutions~II and~III which are also outside the range of 
the CKM fit. The solution~I with $r_c = 0.51 \pm 0.10$,
$\delta_c =(- 39 \pm 10)^\circ$, and 
$\gamma = (65 \pm 5)^\circ$ is the only one which 
agrees well with the global CKM fitted value of~$\gamma$.

The graphical representation of all the solutions 
from Table~\ref{tab:fit-results} is shown in 
Fig.~\ref{fig:isospin}. If one introduces the 
angle~$\theta$ between~${\cal A}^{+-}$ 
and~$\tilde {\cal A}^{+-} = {\rm e}^{+ 2i\gamma} \, 
\bar {\cal A}^{+-}$ and a similar one~$\theta_n$ 
between~${\cal A}^{00}$ and~$\tilde {\cal A}^{00}$, 
the present analysis leads to the following values: 
\begin{equation} 
|\theta| = 19^\circ, 
\qquad 
|\theta_n| = 26^\circ,
\label{eq:theta-fit}
\end{equation}
for all the best-fit solutions in Table~\ref{tab:fit-results}.

The result~(\ref{eq:theta-fit}) for~$\theta$ can be compared 
with the limits on this angle based on the isospin symmetry 
of the $B \to \pi\pi$ decay amplitudes. Several bounds on 
it are known at present: by Grossman and Quinn 
(GQ)\cite{Grossman:1997jr}, Charles (Ch)\cite{Charles:1998qx}, 
and Gronau et al. (GLSS)\cite{Gronau:2001ff}, and the 
corresponding bounds based on the current experimental data 
are presented in Table~\ref{tab:isospin-bounds}.
\begin{table}[tb]
\caption{The isospin-based conservative bounds on the 
         angle~$\theta$.} 
\label{tab:isospin-bounds}
\begin{center} 
\begin{tabular}{|c|cc|} 
\hline  
Bound & 
$\cos(2\theta)_{\rm min}^{\rm cons}$ & 
$|\theta_{\rm max}^{\rm cons}|$  
\\ \hline 
GQ & 0.27 & $37.1^\circ$ \\ 
Ch & 0.29 & $36.6^\circ$ \\ 
GLSS & 0.31 & $36.0^\circ$ 
\\ \hline  
\end{tabular}
\end{center}
\end{table}
%

\begin{figure}[tb] 
\centerline{
\includegraphics[width=0.5\textwidth]{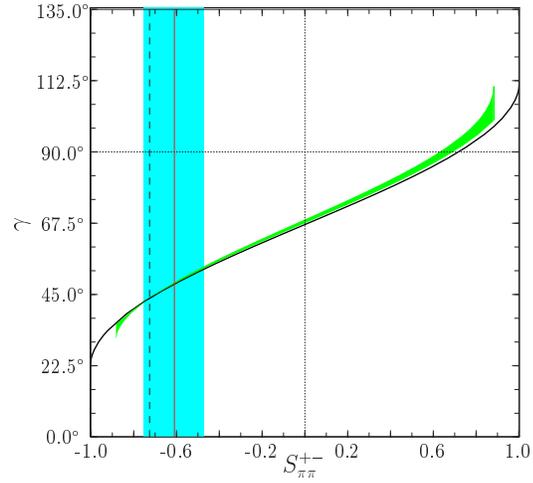}
}
\caption{The bounds on the UT angle~$\gamma$ resulting from the 
         recent data on the CP asymmetry in the $B^0 \to \pi^+\pi^-$ 
         decay. The shaded vertical band is the recent experimental 
         range of the mixing-induced CP asymmetry~$S_{\pi\pi}^{+-}$. 
         The dashed vertical line shows the position of the point 
         $S_{\pi\pi}^{+-} = - \sin(2\beta)$ starting from which 
         the lower limit on~$\gamma$ can be applied.}
\label{fig:Spipi-gamma}
\end{figure}

In the papers\cite{Buchalla:2003jr,Botella:2003xp} a new 
bound on the angle~$\gamma$ from the time-dependent 
CP asymmetry in the $B^0 \to \pi^+ \pi^-$ decay was derived. 
Note that a lower or upper limit results in dependence on 
either $S_{\pi\pi}^{+-} > - \sin (2\beta)$ or 
$S_{\pi\pi}^{+-} < - \sin (2\beta)$. 
As seen from Fig.~\ref{fig:Spipi-gamma}, the value 
$S_{\pi\pi}^{+-} = - \sin (2\beta)$ (the dashed vertical  
line) is inside the present experimental interval 
of~$S_{\pi\pi}^{+-}$ (the shaded vertical band) and, hence, 
this bound can not be applied at present to constrain $\gamma$.
%

\section{Summary}
\label{sec:summary}

Present data on the $B \to \pi\pi$ decays allow to perform 
a phenomenological isospin amplitude analysis in the limit 
of neglecting the electroweak-penguin contributions without 
any additional restriction on the UT angle~$\gamma$. 
The fitting procedure results four best-fit solutions 
from which only one with $\gamma = (65 \pm 5)^\circ$ is 
in agreement with the global CKM fit of the SM. The ratio 
of the penguin-to-tree contributions $r_c = 0.51 \pm 0.10$
is now smaller in comparison with the previously derived 
value\cite{Ali:2004hb} ($r_c = 0.77^{+0.58}_{-0.34}$) 
and has moved in the direction of the typical predictions 
of the dynamical models $r_c \simeq 0.30$. The phase of this 
ratio $\delta_c = (- 39 \pm 10)^\circ$ can be well understood 
in the framework of the pQCD approach\cite{Keum:2002vi} 
while it remains problematic within the 
QCD-Factorization\cite{Buchalla:2003jr}. 
Recent data support the abnormally enhanced 
color-suppressed contribution to the decay amplitudes:
$x_c = 1.11 \pm 0.10$ and $\Delta_c = (-56 \pm 13)^\circ$, 
which are well correlated with the previous 
estimates\cite{Buras:2003dj}: $x_c = 1.22^{+0.26}_{-0.21}$
and $\Delta_c = (-71^{+19}_{-26})^\circ$, obtained by 
applying the flavor $SU(3)$ symmetry to the $B \to \pi\pi$
and $B \to K \pi$ modes\footnote{After the completion of our 
analysis, the paper\cite{Buras:2004th} appeared in which the 
updated values $x_c = 1.13^{+0.17}_{-0.16}$ and 
$\Delta_c = (-57^{+20}_{-30})^\circ$ have been obtained. 
They agree well with our best-fit solution~I.}. 
Such an enhanced color-suppressed 
contribution can not be successfully explained within the 
current theoretical approaches, except possibly SCET assuming 
dominance of the charming penguins.
As for the electroweak-penguin contributions which were 
neglected in the present analysis but can be of potential 
importance\cite{Buras:2003dj}, recent theoretical analyses 
and updated experimental data\cite{Charng:2004ed,Nandi:2004dx} 
reduce their size in decay amplitudes, decreasing the  
possibility of discovering new physics in charmless 
two-body $B$-meson decays.

\section*{Acknowledgements}

This work was supported in part by the Schweizerischer
Nationalfonds and by the Council on Grants by the President 
of Russian Federation for the Support of Young Russian 
Scientists and Leading Scientific Schools of Russian 
Federation under the Grant No. NSh-1916.2003.2. 
A.P. would like to thank Ahmed Ali for discussions and 
the DESY Theory Group for the hospitality where a part 
of the work was done.

\end{document}